\documentclass[final,5p,times,twocolumn]{elsarticle}
\biboptions{sort&compress,square,comma,numbers}
\usepackage{amssymb}
\usepackage{lipsum}
\usepackage{graphicx}
\usepackage{amsmath}
\usepackage{slashed}
\usepackage{amsfonts}
\usepackage{amssymb}
\usepackage{parskip}
\usepackage{xcolor,soul}
\usepackage{colortbl}
\usepackage{epstopdf}
\usepackage{xcolor}
\usepackage{float}
\usepackage{subfigure}
\usepackage{hyperref}
\usepackage[hyphenbreaks]{breakurl}
\usepackage{relsize}
\usepackage{graphics}
\usepackage{hyperref}
\usepackage{mathrsfs}
\usepackage{amssymb}
\usepackage{booktabs}
\usepackage[normalem]{ulem}
\usepackage{amsthm}
\usepackage{cancel}
\usepackage{natbib}
\usepackage{orcidlink}
\usepackage{tcolorbox}
\usepackage{tabularx,ragged2e} 
\usepackage[nodisplayskipstretch]{setspace}
\usepackage{tikz}
\usepackage[compat=1.1.0]{tikz-feynman}
\usepackage{multirow}
\usepackage[font=small]{caption}
\usepackage[font={small},labelfont={bf}]{caption}
\usepackage{url}
\definecolor{MyDarkBlue}{rgb}{0.1, 0.1, 0.8}
\definecolor{SBlue}{rgb}{0.2, 0.4, 0.7} 
\definecolor{MyLightBlue}{rgb}{0.22,0.51,0.9}
\definecolor{MyGreen}{rgb}{0.0, 0.5, 0.0}
\definecolor{BrickRed}{rgb}{0.8, 0.25, 0.33}
\usepackage{hyperref}
\hypersetup{colorlinks, citecolor=BrickRed,linkcolor=MyDarkBlue, urlcolor=MyGreen}
\usepackage{physics}
\usepackage{tikz}

\newcommand{\mdm}{m_\text{DM}}
\newcommand{\Trh}{T_\text{RH}}
\newcommand{\Tmax}{T_\text{max}}
\newcommand{\gs}{g_\star}
\newcommand{\gss}{g_{\star s}}

\date{}
\journal{Physics Letters B}
\begin{document}
\begin{frontmatter}
\title{Lepton collider as a window to reheating via freezing in dark matter detection. Part I}
\author[a]{Basabendu Barman\orcidlink{0000-0003-0374-7655}}
\ead{basabendu.b@srmap.edu.in}
\affiliation[a]{Department of Physics, School of Engineering and Sciences, SRM University-AP, Amaravati 522240, India}
\author[b]{Subhaditya Bhattacharya\orcidlink{0000-0002-8841-603X}}
\ead{subhab@iitg.ac.in}
\author[b]{Sahabub Jahedi\orcidlink{0000-0003-1016-8264}}
\ead{sahabub@iitg.ac.in}
\author[b]{Dipankar Pradhan\orcidlink{0000-0002-2450-6677}}
\ead{d.pradhan@iitg.ac.in}
\author[b]{Abhik Sarkar\orcidlink{0000-0003-1449-2934}}
\ead{sarkar.abhik@iitg.ac.in}
\affiliation[b]{Department of Physics, Indian Institute of Technology, Guwahati, Assam-781039, India}
\begin{abstract}
{We propose a methodology to infer the reheat temperature ($\Trh$) of the Universe from the collider signal of freezing in dark matter (DM). We demonstrate it for the mono-$\gamma$ signal at the electron-positron colliders, which indicates to a low-scale $\Trh$, after addressing observed DM abundance, BBN, and other relevant constraints. The method can be used to correlate different reheating dynamics, DM models, and collider signals.}
\end{abstract}
\end{frontmatter}
High energy particle colliders are essential tools for new physics (NP) search, where we can create heavy particles close or below the available center-of-mass (CM) energy. Our study attempts to address whether a correspondence can be established between collider signal and the early Universe cosmology, focusing particularly on the pre-Big Bang Nucleosynthesis (BBN) epoch. Through this approach, we can connect a large class of Dark Matter (DM) models and signals. Few attempts have been made before to probe FIMP DM at the colliders assuming the presence of a charged particle in the dark sector~\cite{Belanger:2018sti,Becker:2023tvd}. In this work, we consider the direct production of FIMP DM at the colliders corresponding to low temperature reheating.

The cosmology of the early Universe has a profound impact on the production of exotic relics, for example, the DM, that constitutes about $26\%$ of the matter-energy budget of the Universe~\cite{Planck:2018vyg}. Many questions revolve around DM: what is its fundamental nature, what is the production mechanism, or, how to possibly detect such entities. A particle DM is well motivated to fit the observations, having any intrinsic spin, minimal photon coupling, and stable of the order of Universe's life time. Depending on whether they were produced in equilibrium with the thermal bath or not, they are classified mainly as weakly interacting massive particles (WIMPs)~\cite{Jungman:1995df} and feebly interacting massive particles (FIMPs)~\cite{Hall:2009bx}. Beyond that, there are other DM productions which include axion misalignment \cite{Preskill:1982cy,Dine:1982ah}, production in inflaton decay \cite{Moroi:2020has}, gravitational production \cite{Ford:1986sy}, etc.

FIMPs, having feeble interaction with the visible Universe, 
can be produced from out-of-equilibrium decay or scattering of the SM particles and can easily evade stringent experimental bounds, unlike WIMPs. In a typical UV freeze-in~\cite{Elahi:2014fsa}, DM yield becomes proportional to the temperature of the thermal bath~\cite{Elahi:2014fsa}. Thus bulk of the DM is produced at the highest temperature of the Universe, usually known as the reheating temperature, when reheating happens instantaneously. The crucial point is, since freeze-in occurs out of equilibrium, DM retains the ``memory" of the early Universe cosmology. Detecting such relics at the colliders could, therefore, provide a probe of the reheating temperature or early Universe cosmology. 

The Effective Field Theory (EFT) approach~\cite{Goodman:2010ku,Beltran:2008xg,Cao:2009uw,Fedderke:2014wda,Liang:2023yta} to search for DM is economic in parameters and efficient in capturing the essential characteristics, focusing solely on the relevant degrees of freedom present at a certain scale. EFT methodologies are widely used, including condensed matter systems (see,~\cite{Brauner:2022rvf}), and have emerged as a standard prescription for studying physics beyond the SM, but limited, if collider experiments reveal a direct manifestation of the heavy mediator, generating effective interactions. EFT is therefore valid when the CM energy of the collision remains lower than the heavy mediator mass. This is easier to achieve in lepton colliders than hadronic ones, since the partonic CM energy differs from the exact CM energy in the latter case.

At the colliders, as DM evades detection, the visible particles produced from the initial state radiation (ISR) in association with DM can provide vital clues via momentum or energy imbalance to search for DM. Mono-$X$, ($X\in\gamma,\,j,\, Z,\, H$) plus missing energy ($\slashed{E}$) final state is one of the popular probes to search for DM at colliders. The signal of our interest is mono-$\gamma$, which has been studied extensively in lepton colliders~\cite{Fox:2011fx, Yu:2013aca, Essig:2013vha, Kadota:2014mea, Yu:2014ula, Freitas:2014jla, Dutta:2017ljq, Choudhury:2019sxt, Horigome:2021qof, Barman:2021hhg, Kundu:2021cmo, Bhattacharya:2022qck,Ge:2023wye,Ma:2022cto,Singh:2024wdn,Borah:2024twm,Fayet:1982ky,Ellis:1982zz,Grassie:1983kq}. Noticeably, the final state with ISR photon has similar $\slashed{E}$ event distribution to the huge SM neutrino background ($\nu\bar{\nu}\gamma$), making it difficult for signal-background separation (see \ref{sec:app-radvsver}). This motivates us to look for associated photon production, which we call {\em natural} mono-$\gamma$ signal within the EFT framework. 

Dimension-six and seven EFT operators with real scalar $(\Phi)$ and fermionic $(\chi)$ DM leading to natural mono-photon searches are\footnote{Other mono-X channels involving a different set of operators, or opposite-sign electron signal can also explain this connection, see \ref{sec:app-radvsver} and \cite{Barman:2024tjt}.},
\begin{align}     &\mathcal{O}_{2}^{s}:~\dfrac{c_{\Phi}}{\Lambda^2}(B_{\mu\nu}B^{\mu\nu}+W_{\mu\nu}^i\,W^{i\mu\nu})\,\Phi^2\,,\label{eq:os2}
\\&     \mathcal{O}_{3}^{f}:~\dfrac{c_{\chi}}{\Lambda^3}(B_{\mu\nu}B^{\mu\nu}+W_{\mu\nu}^i\,W^{i\mu\nu})\,(\overline{\chi}\chi)\,,\label{eq:of3}
\end{align}
where $W_{\mu \nu}^i$ $(i=1,\,2,\,3)$ and $B_{\mu \nu}$ are the electroweak field strength tensors corresponding to $SU(2)_L$ and $U(1)_Y$ respectively, and $c_{\Phi,\chi}$ are the dimensionless Wilson-coefficients\footnote{It is straightforward to extend the operators involving complex scalar and vector DM leading to natural mono-$\gamma$, as done in \cite{Barman:2024tjt}.}. Unless otherwise stated, we consider $c_{\Phi, \chi}=1$ throughout this analysis\footnote{{\color{black} We assume that 
the operator relevant for DM production at early and present Universe is $\mathcal{O}_{3}^{f}$ for fermionic DM and $\mathcal{O}_{2}^{s}$ for scalar DM, and all the operators having lower mass dimension is absent. This makes the Higgs portal interaction $\Phi^2 |H|^2$ to be zero as well.}}.
The presence of DM fields in pair implies $\mathcal{Z}_2$-symmetry that ensures absolute stability of the DM. These operators are well studied in the context of freeze-out, with a primary focus to constrain direct and indirect searches, complementing the LHC search bounds~\cite{PhysRevD.89.056011,Arina:2020mxo,Kavanagh:2018xeh}.
\begin{figure*}[htb!]    
\includegraphics[width=0.45\linewidth]{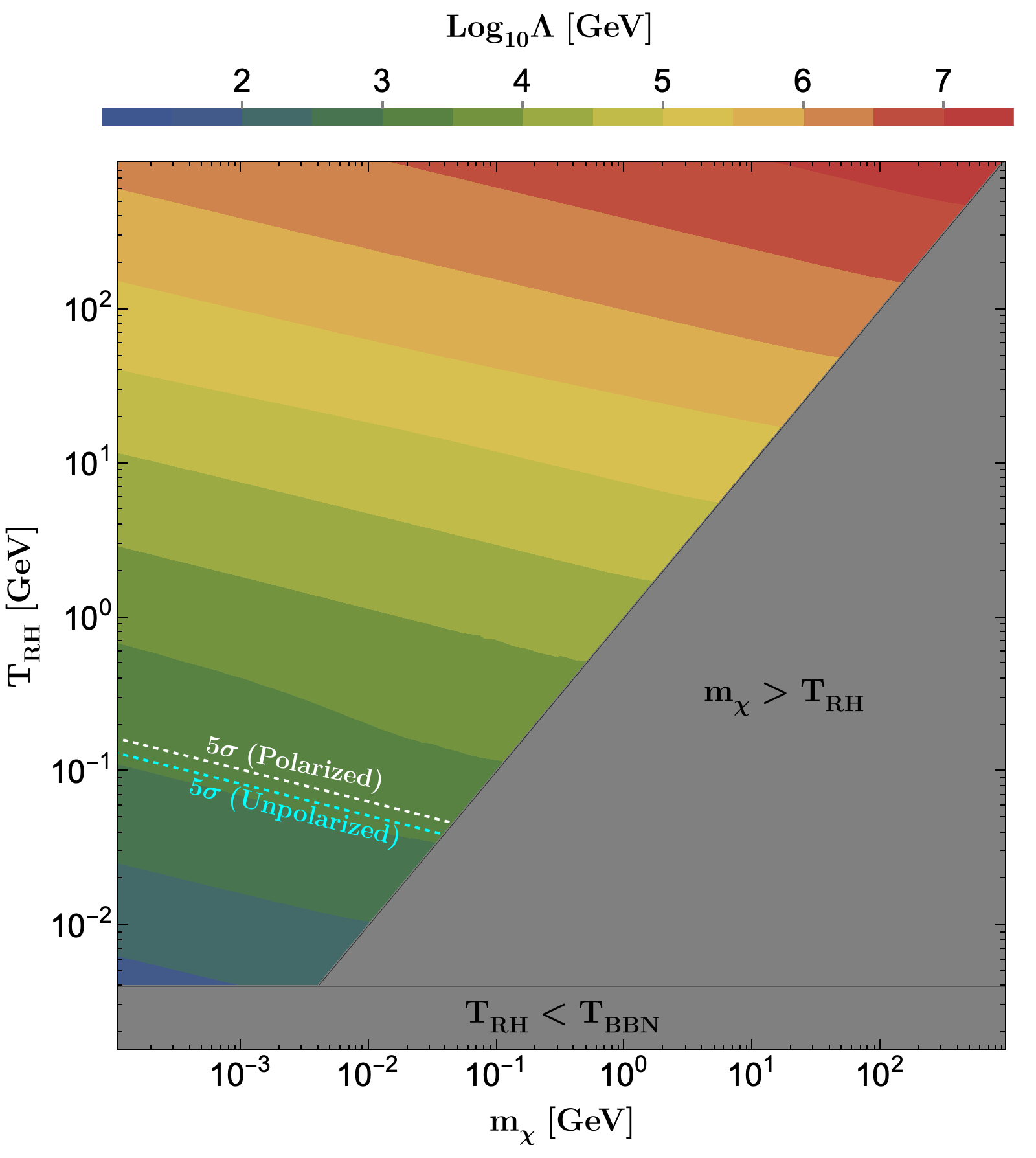}\quad
\includegraphics[width=0.45\linewidth]{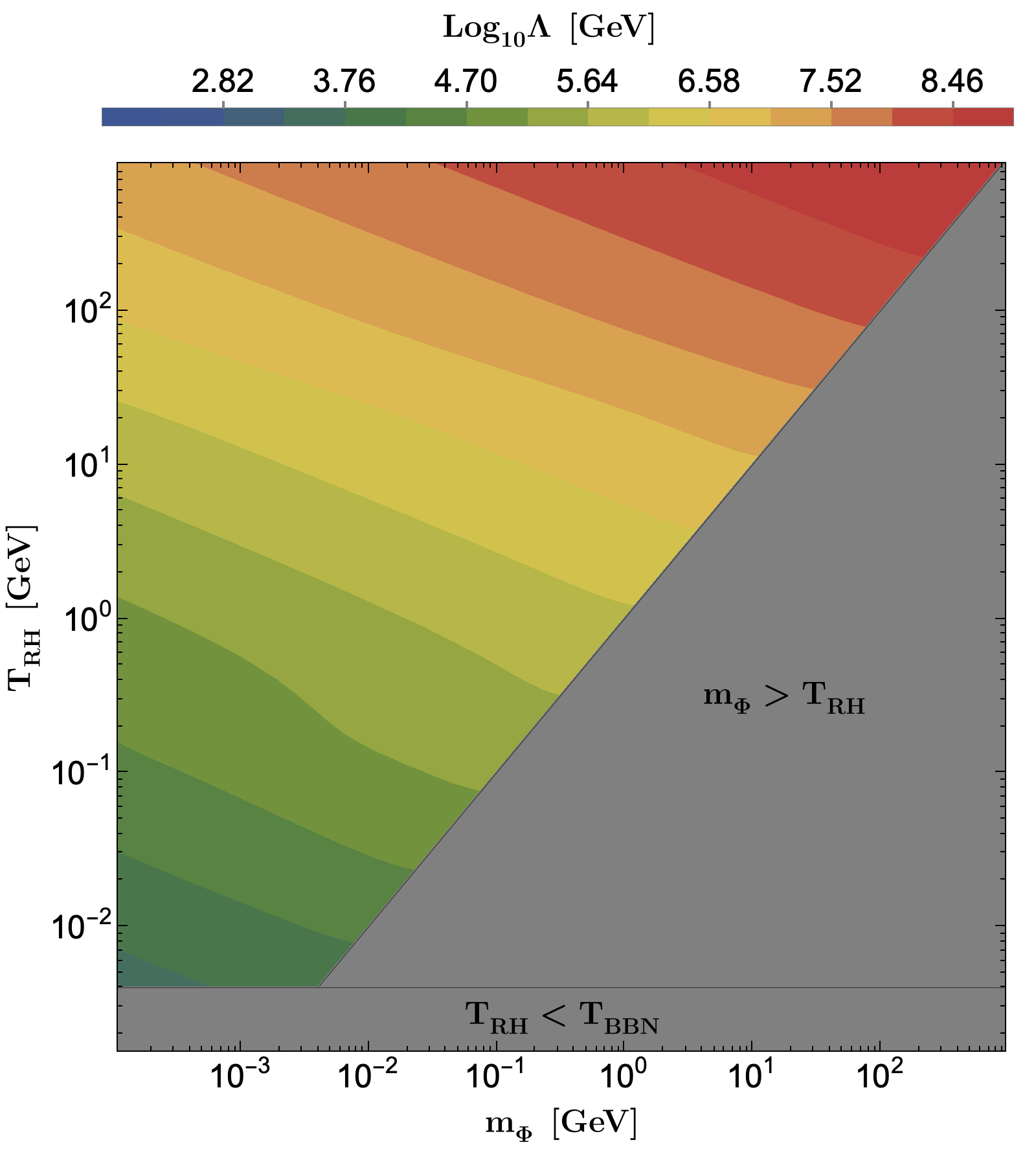}
\caption{DM relic allowed parameter space for fermion (left) and scalar (right) DM, where the colour bar corresponds to $\Lambda$. The gray shaded regions are disallowed from the instantaneous decay approximation requiring $\mdm\lesssim\Trh$, as well as from the BBN bound $\Trh\gtrsim 4$ MeV. In the left panel the cyan and white dashed contours shows 5$\sigma$ significance for mono-$\gamma+\slashed{E}$ signal at the ILC, with $\sqrt{s}=$ 1 TeV, $\mathfrak{L}_{\text{int}}$ = 8 $\rm{ab^{-1}}$ for unpolarized and polarized ($\{P_{e^+},P_{e^-}\}=\{-20\%,+80\%\}$) beams respectively.}
\label{fig:fimp-scan}
\end{figure*}
We discuss DM production via freeze-in and derive the viable parameter space that complies with the observed DM abundance. The essential feature of freeze-in production is to consider that the DM number density was vanishingly small in the early Universe, then produced dynamically from the visible sector via DM-SM interactions, $\gamma\gamma\to$ DM\,DM in our case, emerging from Eq.~\eqref{eq:os2} and \eqref{eq:of3}. The DM-SM interaction strength is several orders of magnitude feebler than the weak interaction strength, so that the DM stays out of equilibrium. The limiting condition can be estimated by comparing the DM interaction rate $\Gamma_{2\to2}=n_{\rm DM}^{\rm eq}\,\langle\sigma v\rangle_{{\rm DM\,DM}\to\gamma\gamma}$ with the Hubble rate $H=\pi/3\,\sqrt{\gs/10}\,T^2/M_P$, where $\langle\sigma v\rangle$ is the thermally averaged DM production cross-section~\cite{Gondolo:1990dk},
\begin{align}
& \frac{\Gamma_{2\to2}}{H}\Bigg|_{T=\Trh}\simeq\frac{1}{\sqrt{\gs(T)}}
\nonumber\\&\times
\begin{cases}
\mathcal{C}_0\,\left(\frac{m_\Phi}{{1}\,{\rm MeV}}\right)\left(\frac{\Trh}{{4}\,{\rm MeV}}\right)^2\,\left(\frac{{1}\,{\rm TeV}}{\Lambda}\right)^4
& \text{for scalar}\,,
\\[10pt]
\mathcal{C}_{1/2}\,\left(\frac{m_\chi}{1\,{\rm MeV}}\right)^3\left(\frac{\Trh}{{4}\,{\rm MeV}}\right)^2\,\left(\frac{1\,{\rm TeV}}{\Lambda}\right)^6 & \text{for fermion}\,,
\end{cases}
\end{align} 
Here, $\mathcal{C}_0\simeq{1.39\times 10^{-1}}$ and $\mathcal{C}_{1/2}\simeq{8.94\times 10^{-16}}$, achieved by a conservative estimate of assuming DM number density to be the same as the equilibrium number density, i.e. $n_{\rm DM}\equiv n_{\rm DM}^{\rm eq}=T/(2\,\pi^2)\,\mdm^2\,K_2\left(\mdm/T\right)$. As one can infer, it is easy to satisfy the out-of-equilibrium condition for fermionic DM compared to the scalar case due to larger suppression (see \autoref{tab:rate-DM} of \ref{sec:app-reacden}). The very idea that we wish to investigate the collider signature of freeze-in, drives us to choose $\Lambda\sim\mathcal{O}$(TeV) such that the collider signal is significant enough, which in turn demands $\mdm\sim\mathcal{O}$(MeV) to avoid overclosing the Universe. 

Because of the non-renormalizable nature of the interaction, we see, DM yield features the typical UV nature~\cite{Elahi:2014fsa} (see \ref{sec:app-reacden}), where bulk of the DM is produced around the highest temperature. 
For DM production via $\gamma\,\gamma\to$ DM\,DM channel, an approximate analytical expression for DM yield $Y=n/s$ can be obtained by solving the corresponding Boltzmann equation (BEQ) (see \ref{sec:app-reacden})
\begin{align}\label{eq:DM-yield}
& Y_{\rm DM}(T)=\frac{270\,\sqrt{10}\,M_P}{\gss(T)\,\sqrt{\gs(T)}\,\pi^8}
\begin{cases}
\frac{\Trh^3-T^3}{\Lambda^4} & \text{for scalar} \,,
\\[10pt]
\frac{288}{5}\,\frac{\Trh^5-T^5}{\Lambda^6} & \text{for fermion}\,,
\end{cases}
\end{align}
considering $s\gg 4\,\mdm^2${\color{black}, where $\gs$ and $\gss$ are the number of relativistic degrees of freedom (DOF) contributing to energy and entropy density, respectively}. Here, $\Trh$ is the reheating temperature which, in the approximation of a sudden decay of the inflaton, corresponds to the maximal temperature reached by the SM thermal bath. Away from the sudden decay approximation for reheating, the bath temperature may rise to a temperature $\Tmax\gg\Trh$~\cite{Giudice:2000ex}. It is plausible that the DM relic density may be established during this reheating period, in which case its abundance will significantly differ from freeze-in calculations assuming radiation domination. However, as the precise determination of such phenomenon depends on the details of the reheating mechanism (in particular, the shape of the inflationary potential during reheating), it is beyond the scope of the present study.

To fit the observed DM relic density, 
$Y_0\, \mdm = \Omega_{\rm DM} h^2 \, \frac{1}{s_0}\,\frac{\rho_c}{h^2} \simeq 4.3 \times 10^{-10}~\text{GeV}\,,$ where $Y_0$, is the present day yield. Here $\rho_c \simeq 1.05 \times 10^{-5}\, h^2$~GeV/cm$^3$ is the critical energy density, $s_0\simeq 2.69 \times 10^3$~cm$^{-3}$ the present entropy  density~\cite{ParticleDataGroup:2022pth}, and $\Omega h^2 \simeq 0.12$ is the observed abundance of DM relics~\cite{Planck:2018vyg}. From Eq.~\eqref{eq:DM-yield}, it is clear that, for a given DM mass and effective scale, the yield is maximum at $\Trh\gg T$, which is the quintessential feature of UV freeze-in. The DM abundance is governed by three independent parameters: $\{\mdm,\,\Trh,\,\Lambda\}$. The effective description of Eq.~\eqref{eq:os2} and \eqref{eq:of3} remains valid under the hierarchy: $\Lambda > \Trh\gtrsim \mdm$, implying, the cut-off scale $\Lambda$ stands as the highest scale of the theory {\color{black}(detailed in \ref{app:EFT})}. In case of instantaneous decay approximation, maximum mass for the DM produced from the thermal bath can be $\mdm\simeq\Trh$. In order to produce the observed DM abundance, using eq.~\eqref{eq:DM-yield} we find
\begin{align}
& \Lambda\simeq 
\begin{cases}
5\,{\rm TeV}\,\left(\frac{m_\Phi}{1\,{\rm MeV}}\right)^{1/4}\,\left(\frac{\Trh}{T_{\rm BBN}}\right)^{3/4} & \text{for scalar}\,,
\\[10pt]
100\,{\rm GeV}\,\left(\frac{m_\chi}{1\,{\rm MeV}}\right)^{1/6}\,\left(\frac{\Trh}{T_{\rm BBN}}\right)^{5/6} & \text{for fermion}\,,
\end{cases}
\end{align}
which shows, in order to have $\Lambda\simeq\mathcal{O}(\rm TeV)$, we need to opt for low DM mass, as well as low $\Trh$. It is easier to achieve for $\mathcal{O}_3^f$, because of larger suppression. {\color{black} We emphasize that the above estimation is approximate, as it accounts only for $\gamma\,\gamma\to$ DM,\,DM as the main production channel, and considers a constant $\gs\approx\gss\simeq 3$ at $\Trh=T_{\rm BBN}$. A complete numerical solution of the BEQ, including all possible DM production channels together with temperature evolution of the DOFs, has been performed in reality. The approximate estimation merely demonstrates that $\mathcal{O}_3^f$ is a favorable choice in the present scenario, and highlights the pronounced sensitivity of the DM abundance to the maximum temperature of the Universe, a quintessential feature of UV freeze-in.} Note $T_{\rm BBN}\simeq 4$ MeV~\cite{Hannestad:2004px} corresponds to the lower bound on $\Trh$, such that the accurate measurement of light element abundance during BBN is not hampered\footnote{{Since the DM-SM interaction is feeble here, the DM parameter space is safe from the CMB as well as INTEGRAL limit~\cite{PhysRevD.101.123514,Siegert:2024hmr}, that constraints DM annihilation to photon final states.}}. This order of magnitude estimation conforms well with the full numerical results as illustrated in \autoref{fig:fimp-scan}, showing relic density allowed parameter space in $\Trh-m_{\rm DM}$ plane, where different shades of bands correspond to different values of $\Lambda$.
{\color{black}A non-linear behavior in the DM relic-allowed region emerges in the color band around $\Trh\sim160$ MeV, driven by changes in the effective degrees of freedom of the SM bath across the QCD phase transition.
}
Interestingly, 5$\sigma$ significance for the mono-$\gamma+\slashed {E}$ signal produced by the DM at the ILC (with $\sqrt{s}$ = 1 TeV, and integrated luminosity $\mathfrak{L}_{\text{int}}$ = 8 $\rm{ab^{-1}}$) can be projected in this plane, as shown by the (cyan) white dashed lines for (un)polarized beams in the left panel plot, indicating the possibility of collider reach to probe the MeV-scale reheating era. We will elaborate on it soon.

It is worth reminding that MeV-scale reheating is realizable in a minimal gravitational reheating scenario, where the SM bath is produced from inflaton scattering, mediated by graviton~\cite{PhysRevD.107.043531}. Since gravity interacts with all matter particles democratically, such a reheating process is inevitable and natural. The minimal scenario is however dismissed due to the excessive production of inflationary gravitational wave energy density around the time of BBN~\cite{Figueroa:2018twl,Opferkuch:2019zbd}. As mentioned before, we refrain from adhering to a specific reheating mechanism here, and simply consider $\Trh$ as a free parameter. 
\begin{figure}[htb!]
\centering
\includegraphics[width=0.95\linewidth]{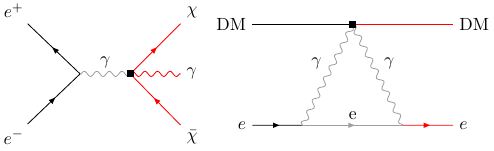}
\caption{{\it Left:} Feynman graph for associated production of fermionic DM pair ($\chi \overline{\chi}$) with mono-$\gamma$ at the $e^{+} e^{-}$ collider. {\it Right:} Feynman diagram corresponds to the FIMP-electron $(\rm DM-e)$ scattering.}
\label{fig:ldm-cd}
\end{figure}
\begin{figure*}[htb!]
\centering   
\includegraphics[width=0.95\linewidth]{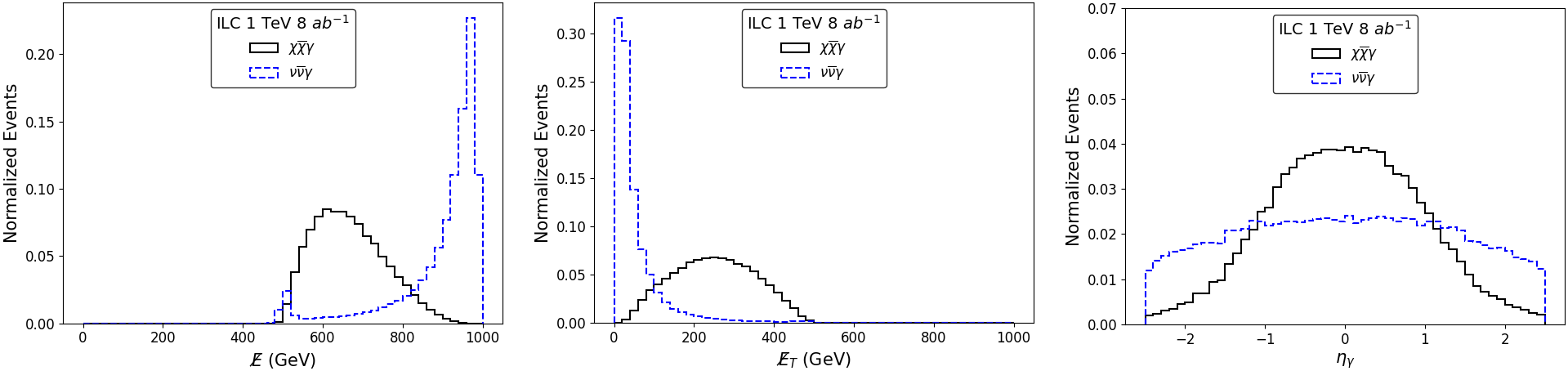}
\caption{{Normalised} signal background event distributions for mono-$\gamma$ final state. Left: ME ($\slashed{E}$), middle: MET ($\slashed{E}_T$), right: Pseudorapidity ($\eta_{\gamma}$). The signal corresponds to: $m_{\chi}$=33 MeV, $\Lambda$=1.14 TeV.}
\label{fig:dist1}
\end{figure*}
\begin{figure*}[htb!]
\centering
\includegraphics[width=0.95\linewidth]{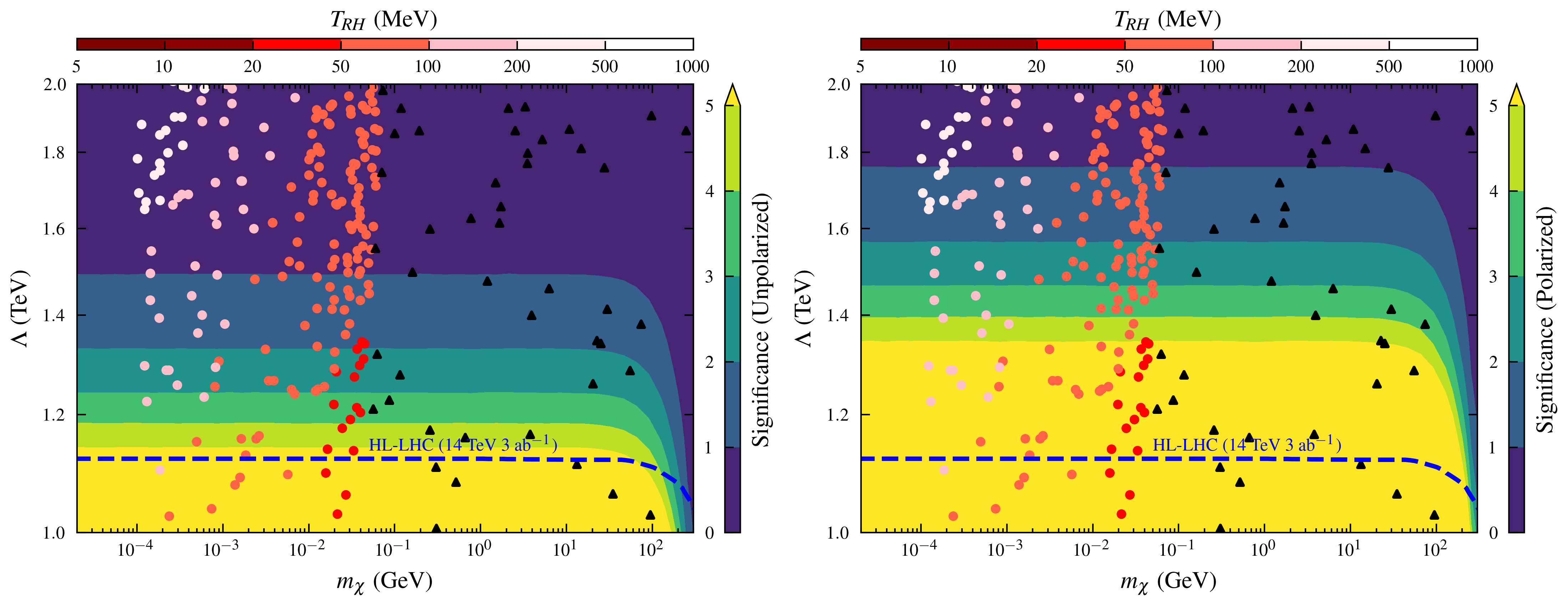} 
\caption{Observed DM abundance for $\mathcal{O}_3^f$ for different combinations of $m_\chi$, $\Lambda$ and $\Trh$, shown via scattered reddish points. Black points are disallowed, requiring $\Trh<T_{\rm BBN}$ or $m_\chi>\Trh$ or both to produce the right abundance; in conflict with the BBN bound and instantaneous reheating approximation. The colour gradient shows the variation of $\mathcal{S}$ at ILC with $\sqrt{s}=$ 1 TeV, and $\mathfrak{L}_{\rm int}=$ 8 $\rm{ab^{-1}}$, for unpolarized beams (left), and polarised beams $\{P_{e^+}, P_{e^-}\}$ = $\{-20\%, +80\%\}$ (right). The blue dashed line represents the projected 95\% C.L. exclusion limit from the HL-LHC ($\sqrt{s}=14$ TeV, $\mathfrak{L}_{\rm int}=3$ $\rm{ab^{-1}}$) \cite{Bruning:2015dfu}.}
\label{fig:final1}
\end{figure*}
The photophilic DM provides mono-$\gamma$ with missing energy signal as the innate choice, as shown in the left panel of \autoref{fig:ldm-cd}. For collider simulation we consider $\mathcal{O}_3^f$, since for $\mathcal{O}_2^s$, one needs $\Trh<T_{\rm BBN}$ to have $\Lambda\lesssim\mathcal{O}$(TeV), as seen from \autoref{fig:fimp-scan}. The choice of $\Lambda$ is consistent with the LEP {and the LHC mono-photon data} (see \ref{app:LEP} and \ref{app:LHC}). For larger $\Lambda$, the signal significance diminishes as the production cross-section reduces. We perform detector-level analysis at the projected $\sqrt{s}=1$ TeV and $\mathfrak{L}_{\text{int}}=8~\rm{ab^{-1}}$ which is expected to be the maximum reach of the ILC~\cite{ILCInternationalDevelopmentTeam:2022izu}. The model implementation is done in \texttt{FeynRules}~\cite{Alloul:2013bka} and the signal/background events are generated using \texttt{MG5\_aMC}~\cite{Alwall:2011uj}. The simulated events are then showered using \texttt{Pythia8}~\cite{Sjostrand:2014zea} and further detector simulation is done exploiting \texttt{Delphes3} \cite{deFavereau:2013fsa}. 

The signal selection is ensured by only allowing events with a single photon with no associated leptons or jets. The dominant SM background stems from $\nu\bar\nu\gamma$ channel, where the photon can be radiated from the electron leg in the $W$ mediated $t$-channel (dominant), $Z$ mediated $s$-channel or from the $W$ boson fusion process. The photon from the SM neutrino background is characterized by low {\it transverse momentum} $(p_{T})$, unlike the signal. Thus, imposing {\it missing transverse energy} (MET), $\slashed{E}_{T}=-\sqrt{\sum\left(p_x\right)^2+\sum\left(p_y\right)^2}$ (where the sum runs over all visible objects) cut effectively mitigates a substantial portion of the dominant SM background. {\it Missing energy} (ME) in this case is defined as $\slashed{E} = \sqrt{s} - E_{\gamma}$. Applying a cut to ensure the removal of the two peaks of the neutrino $\slashed{E}$ distribution reduces the background significantly, retaining most of the signal. Finally, implementing an absolute {\it pseudorapidity} $|\eta_\gamma|<1.0$ finely tunes the elimination of non-transverse backgrounds. For the benchmarks illustrated in \autoref{tab:sbc}, the signal efficiency\footnote{We define $\epsilon_{s}=\sigma_{\tt sig}/\sigma_{\tt prod}$, where $\sigma_{\tt prod}$ and $\sigma_{\tt sig}$ are the signal cross-section before and after cut, respectively.} $\epsilon_s$ is obtained to be 0.51 for a background rejection rate of 99\%. The event topology in \autoref{fig:dist1} shows the impact of these particular observables in separating the signal from the SM background. For details of simulation techniques, see \ref{app-colldet}.
The signal significance is calculated via~\cite{Cowan:2010js} 
\begin{equation}
\mathcal{S} = \sqrt{2 \left[(S + B) \log \left(1 + \frac{S}{B} \right) - S \right]}\,.
\end{equation}
Here $S$ and $B$ denote signal and background events, respectively. Signal significance ($\mathcal{S}$) undergoes enhancement after employing a series of cuts outlined in \autoref{tab:sbc}. Note that, for $S/B\ll 1$, one finds $\mathcal{S}\simeq S/\sqrt{B}\propto 1/\Lambda^6$.
\begin{table}[htb!]
\centering
\renewcommand{\arraystretch}{1.2}
{\small
\begin{tabular}{|c|c|c|c|}
\hline
\multirow{2}*{Cutflow} & \multirow{2}*{$\nu \overline{\nu} \gamma$ ($B$)} & \multicolumn{2}{c|}{$\chi \overline{\chi} \gamma$ } \\\cline{3-4}
&  & Events ($S$) & Significance ($\mathcal{S}$) \\\hline
Basic Cuts & 18101325 & 4422 & 1.04 \\
$\slashed{E}_T$ $>$ $200.0$ GeV & 725945 & 2957 & 3.47 \\
$\slashed{E}$ $\in$ $[525, 750]$ GeV & 289420 & 2694 & 5.00 \\
$|\eta_{\gamma}| <$  1.0 & 219161 & 2395 & 5.10 \\\hline
\end{tabular}}

\vspace{0.25 cm}

\renewcommand{\arraystretch}{1.2}{
\begin{tabular}{|c|c|c|c|c|c|}
\hline
\multirow{2}{*}{\shortstack[c]{Polarization \\$\{P_{e^{+}}, P_{e^{-}}\}$ }}& \multirow{2}*{$\nu \overline{\nu} \gamma$ ($B$)~} &  \multicolumn{2}{c|}{$\chi \overline{\chi} \gamma$ } \\\cline{3-4}
& & Events ($S$) & Significance ($\mathcal{S}$) \\\hline
$\{0 \%,  0 \%\}$ & 219161 & 2395 & 5.10 \\
$\{+ 20 \%, + 80 \%\}$ & 55052  & 2010 & 8.52 \\
$\{- 20 \%, + 80 \%\}$ & 40711  & 2778 & 13.62 \\
$\{+ 20 \%, - 80 \%\}$ & 453639 & 2774 & 4.11 \\
$\{- 20 \%, - 80 \%\}$ & 303079 & 2003 & 3.63 \\
\hline
\end{tabular}}
\caption{{\it Upper:} Signal-background event counts and $\mathcal{S}$ for mono-$\gamma$ signal for fermion DM, where $m_{\chi}$=33 MeV and $\Lambda$=1.14 TeV. {\it Lower:} Same after final cut for different polarization combinations. {\color{black}The event counts correspond to $\sqrt{s} = 1$ TeV and $\mathfrak{L}_{\rm int} = 8$ ab$^{-1}$.}}
\label{tab:sbc}
\end{table}
As the SM is a left chiral theory, judicious choice of beam polarizations, as $e^+e^-$ machines are equipped with, can suppress the SM background, and enhance $\mathcal{S}$. Owing to the proposed polarizability of ILC, we study $\mathcal{S}$ for various polarization combinations, as outlined in \autoref{tab:sbc}. The detailed event analysis is summarized in \ref{app-colldet}. $\{P_{e^+},\, P_{e^-}\}=\{-20\%,\,+80\%\}$ emerges as the most favorable combination, enhancing the signal while significantly reducing the background. Having larger SM background and absence of polarization, probing the model at the LHC turns difficult.

\autoref{fig:final1} shows $\mathcal{S}$ in $\Lambda-m_\chi$ plane via coloured bands, which decreases with larger $\Lambda$, as the signal cross-section $\sim 1/\Lambda^6$. Also the significance diminishes with larger $m_\chi$ due to limited phase space. Following Eq.~\eqref{eq:DM-yield}, each $\Lambda-m_\chi$ point indicate to a specific $\Trh$ to satisfy the DM abundance, as shown by the reddish points, establishing a correlation between $\mathcal{S}$ and $\Trh$. For higher DM masses, a lower $\Trh$ is required to achieve the correct DM relic, which violates the lower bound on $\Trh$ from BBN or the instantaneous reheating approximation itself (shown by black points). Therefore, it is possible to associate a signal significance of 5$\sigma$ with MeV-scale reheating temperature and MeV-scale frozen in DM. This is what we represent in the left panel of \autoref{fig:fimp-scan}, via the white (cyan) dashed line(s). Note that a significant part of our parameter space survives the projected 95\% C.L. exclusion limit from the HL-LHC 14 TeV 3 ab$^{-1}$ run (blue dashed line).

An effective DM-electron interaction can also be realized at the 1-loop level [cf. right panel of \autoref{fig:ldm-cd}], that leads to DM direct search via electron scattering, which is typically sensitive to MeV-scale DM~\cite{Essig:2011nj}. However, in our case, the corresponding scattering cross section turns out to be $\lesssim 10^{-51}\,\text{cm}^2$, far beyond the reach of present experiments~\cite{DAMIC:2019dcn,SENSEI:2020dpa,PandaX-II:2021nsg,SuperCDMS:2020ymb,CDEX:2022kcd,XENON:2023cxc}.

To summarize, we demonstrate the possibility to infer the reheat temperature of the universe from DM signal at future lepton colliders. The possibility emerges for DMs that undergo UV freeze-in, having one-to-one correspondence with $\Trh$, $\mdm$ and $\Lambda$ to address correct relic abundance. As an example, we elucidate how mono-$\gamma$ signal of an MeV scale photophilic fermionic DM can point towards MeV scale $\Trh$. Such mono-$\gamma$ signal can be segregated from the SM background, when the photon emerges from DM vertex, unlike ISR. The analysis seeds plethora of possibilities, like incorporating the details of the early Universe dynamics \cite{Barman:2024tjt}, different DM operators, and signals which can provide similar inferences.

\section*{Acknowledgments} DP thanks the UGC for senior research fellowship. DP, SJ, and AS would like to thank Cafe Coffee Day, IITG, where the inception of this idea took place. SB acknowledges WHEPP 2024 for discussion on low mass DM. We acknowledge the use of {\tt TikzFeynman} in generating the Feynman graphs. We also acknowledge Rajesh Mondal for useful discussions.
\appendix
\section{Radiation$-\gamma$ vs. Vertex$-\gamma$}
\label{sec:app-radvsver}
\begin{figure*}[htb!]
\centering
\includegraphics[width=0.95\linewidth]{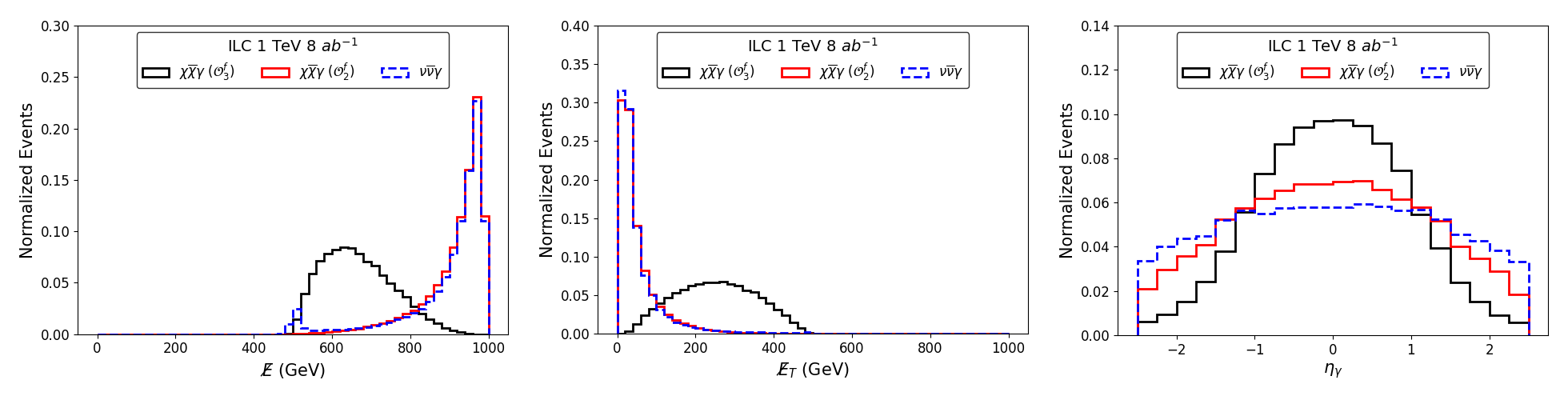}
\caption{Signal background event distributions for different kinematic variables with the mono-$\gamma$ final state signal. Left: MET ($\slashed{E_T}$), middle: ME ($\slashed{E}$), right: Pseudorapidity ($\eta_{\gamma}$). {\color{black}The signal corresponds to: $m_{\chi}$=33 MeV and $\Lambda$=1.14 TeV.}}
\label{fig:dist1A}
\end{figure*}

We review two scenarios in our analysis. The first scenario (already discussed in the main text) involves the operator $\mathcal{O}^f_3$, where the photon from the lepton collider mono-$\gamma$ signal originates directly from the effective vertex itself. In the alternate scenario, the photon is radiated from the incoming electron/positron leg. The later scenario is relevant to DM produced via leptophilic operators. One such operator is given by
\begin{equation}
\mathcal{O}_{2}^{f} = \frac{c_{\chi}}{\Lambda^{2}} (\overline{\ell}_{L} \gamma^{\mu} \ell_{L} + \overline{e}_{R} \gamma^{\mu} e_{R})(\overline{\chi} \gamma_{\mu} \chi)\,.
\end{equation}
We perform a comparative analysis between the vertex photon from $\mathcal{O}_{3}^{f}$ and the radiative photon from $\mathcal{O}_{2}^{f}$ along with the neutrino background. The kinematic distributions are shown in \autoref{fig:dist1A}. It can be clearly seen that the radiative process is always overshadowed by the SM neutrino background, and hence, a clear distinction is extremely difficult using cut-based methods. To segregate these signal processes, we require polarization tuning (to suppress the background) as well as dedicated multivariate techniques.
\section{Reaction density and Boltzmann equation}
\label{sec:app-reacden}
The reaction density corresponding to 2-to-2 processes reads
\begin{align}
\gamma_{22}&=\int\prod_{i=1}^4 d\Pi_i \left(2\pi\right)^4 \delta^{(4)}\biggl(p_a+p_b-p_1-p_2\biggr)\,f_a{^\text{eq}}f_b{^\text{eq}}\left|\mathcal{M}_{a,b\to1,2}\right|^2
\nonumber\\&
=\frac{T}{32\pi^4}\,g_a g_b\,\int_{s_\text{min}}^\infty ds\,\frac{\biggl[\bigl(s-m_a^2-m_b^2\bigr)^2-4m_a^2 m_b^2\biggr]}{\sqrt{s}} \nonumber\\ &\hspace{3cm} \sigma\left(s\right)_{a,b\to1,2}\,K_1\left(\frac{\sqrt{s}}{T}\right)\label{eq:gam-ann}\,,  \end{align}    
with $a,b(1,2)$ as the incoming (outgoing) states and $g_{a,b}$ are corresponding degrees of freedom. Here $f_i{^\text{eq}}\approx\exp^{-E_i/T}$ is the Maxwell-Boltzmann distribution. The Lorentz invariant 2-body phase space is denoted by: $d\Pi_i=\frac{d^3p_i}{\left(2\pi\right)^3 2E_i}$. The amplitude squared (summed over final and averaged over initial states) is denoted by $\left|\mathcal{M}_{a,b\to1,2}\right|^2$ for a particular 2-to-2 scattering process. The lower limit of the integration over $s$ is $s_\text{min}=\text{max}\left[\left(m_a+m_b\right)^2,\left(m_1+m_2\right)^2\right]$.
\begin{figure}[htb!]
\centering    
\includegraphics[width=0.95\linewidth]{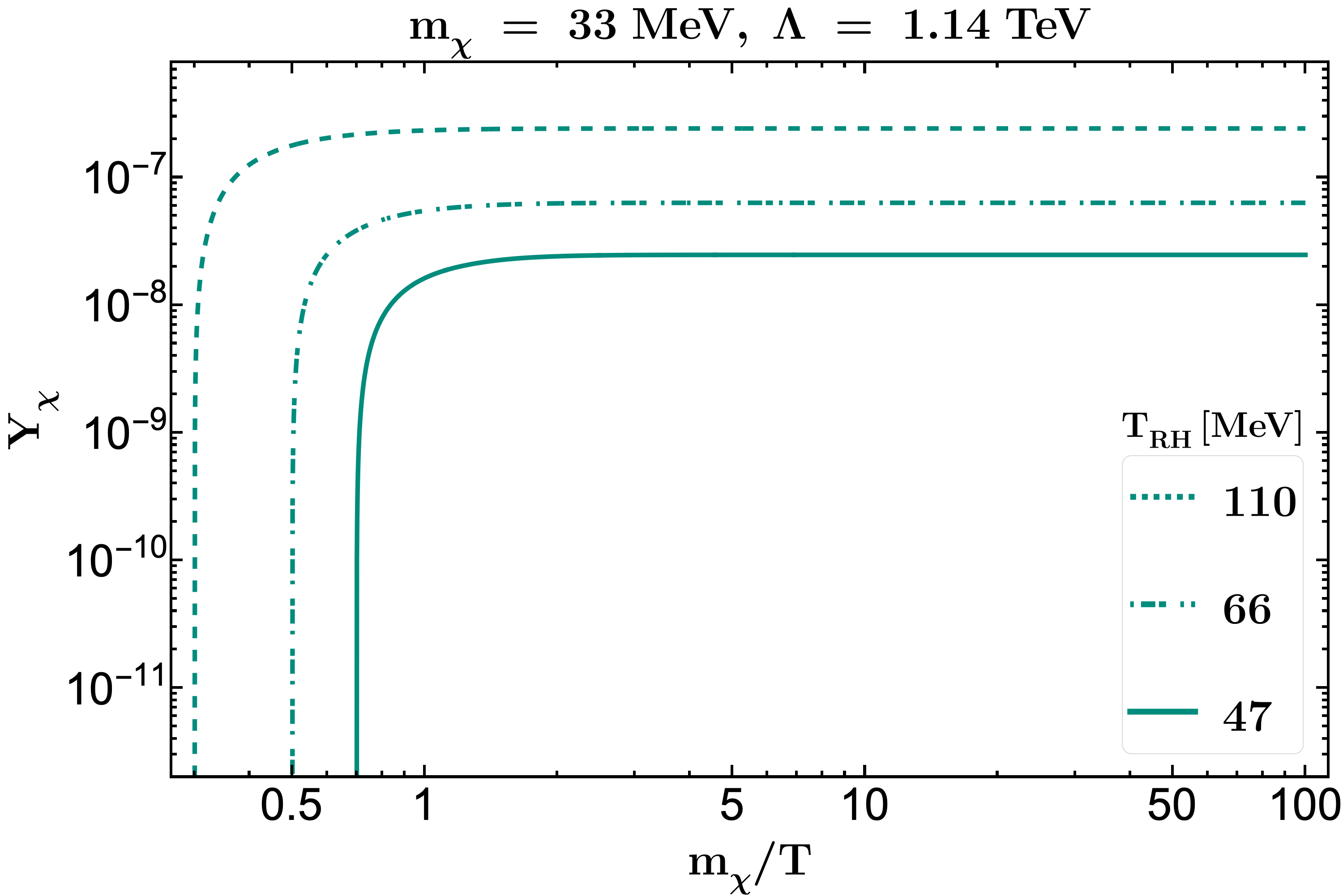}
\caption{Yield of fermionic DM as a function of the dimensionless quantity $m_\chi/T$, where different curves correspond to different choices of $\Trh=\{47,~66,~110\}~\rm MeV$, shown via solid, dot-dashed and dashed patterns, respectively. Here we have fixed $\Lambda=1.14$ TeV and $m_\chi=33$ MeV.}
\label{fig:uv-yield}
\end{figure}

The Boltzmann equation (BEQ) governing the DM number density can be written in terms of the DM yield defined as a ratio of the DM number density to the entropy density in the visible sector, i.e.,  $Y_{\rm DM}=n_{\rm DM}/s$. The BEQ can then be expressed in terms of the reaction densities as
\begin{equation}\label{eq:beq}
x\,H\,s\,\frac{dY_{\rm DM}}{dx} = \gamma_{22}\,,
\end{equation}
where $x\equiv\mdm/T$, $T$ being the temperature of the thermal bath and $H$ is the Hubble parameter. In a radiation dominated Universe, 
\begin{align}
& s(T)=\frac{2\,\pi^2}{45}\,\gss(T)\,T^3\,, & 
H(T)=\frac{\pi}{3}\,\sqrt{\frac{\gs(T)}{10}}\,\frac{T^2}{M_P}\,,
\end{align}
where $\gss$ and $\gs$ are the effective number of relativistic degrees of freedom contributing to the entropy and energy density respectively, while $M_P$ is the reduced Planck mass.
The typical UV nature of the DM yield is very much apparent from \autoref{fig:uv-yield}, where, as one can see, bulk of the DM production happens around $\Trh$, for a given DM mass and NP scale.

In \autoref{tab:rate-DM} we quote the $\Gamma/H$ ratio at $T=\Trh$ for several benchmark values of DM mass, effective scale and reheating temperature, that correspond to the right DM abundance. For all cases we find $\Gamma_{2\to2}/H\ll 1$, showing the DM production happens out of equilibrium (for temperatures lower than $\Trh$, this ratio is even smaller). In each case we also provide the value of the thermally averaged DM,\,DM $\to\gamma\gamma$ cross-section. This shows that our parameter space is safe from CMB bound on the annihilation cross-section of MeV-scale DM into mono-photon final states, that typically requires $\langle\sigma v\rangle\lesssim 10^{-16}\,$ GeV$^{-2}$~\cite{Laha:2020ivk,Siegert:2024hmr}.
\begin{table*}[htb!]
\centering
\setlength{\tabcolsep}{4.5pt} 
\renewcommand{\arraystretch}{1.2} 
\begin{tabular}{|c|c|c|c|c|c|}\hline
DM & $m_{\rm DM}$ (MeV) & $\Lambda$ (TeV) & $ T_{\rm RH}$ (MeV) & $ \Gamma_{2\to 2}/H|_{T=T_{\rm RH}}$ & $\langle \sigma v\rangle_{\rm DM~DM\to \gamma\gamma}$ (GeV$^{-2}$) \\\hline
\multirow{5}{*}{Scalar ($\Phi$)} &0.101&73.621 &439.54 &$2.00\times 10^{-3}$ &$5.02\times 10^{-20}$  \\ \cline{2-6}
&1.097&24.27 &22.44 &$5.52\times 10^{-5}$ &$1.11\times 10^{-20}$  \\ \cline{2-6}
&10.139&68.549 &45.29 &$6.48\times 10^{-6}$ &$7.24\times 10^{-22}$ \\ \cline{2-6}
&101.16&387.26&385.48&$1.80\times 10^{-6}$&$5.19\times 10^{-23}$\\ \cline{2-6}
&1018.6&22233&47863&$6.67\times 10^{-7}$&$7.16\times 10^{-26}$\\ \hline
\multirow{5}{*}{Fermion ($\chi$)}&0.186&1.101&111.69&$1.52\times 10^{-4}$&$3.99\times 10^{-21}$  \\ \cline{2-6}
&1.389&1.076&70.96&$1.99\times 10^{-5}$&$4.28\times 10^{-22}$\\ \cline{2-6}
&12.8&1.248&54&$2.18\times 10^{-6}$&$1.05\times 10^{-22}$\\ \cline{2-6}
&30.6&1.191&43&$9.98\times 10^{-7}$&$6.44\times 10^{-23}$\\ \cline{2-6}
&40&1.204&42&$8.58\times 10^{-7}$&$6.11\times 10^{-23}$\\ \hline
\end{tabular}
\caption{Table showing values of interaction rate to Hubble rate ratio and thermally averaged DM\,DM $\to\gamma\gamma$ cross-sections, at $T=\Trh$, for benchmark points corresponding to right relic abundance. \textcolor{black}{Note that these numbers are evaluated numerically by solving the Boltzmann equation.}}
\label{tab:rate-DM}
\end{table*}
\section{Details of the Collider Analysis}
\label{app-colldet}
In this section, we detail out the event analysis configurations and strategy briefly explained in the letter. A pure mono-$\gamma$ signal is ensured by a no lepton no jet veto. The details of photon, lepton and jet identification and isolation will be detailed later. The signal process is $e^{+} e^{-} \rightarrow \chi \overline{\chi} \gamma$ and event generation is done in \texttt{MG5\_aMC} at the leading order (LO) at center-of-mass (CM) energy of 1 TeV with the following pre-defined cuts on the $\gamma$ kinematics: $p^{\gamma}_{T} > 10$ GeV and $|\eta_{\gamma}| < 2.5$. The dominant background process is $e^{+} e^{-} \rightarrow \nu \overline{\nu} \gamma$. The production level cross sections for the signal and the major background are listed for different CM energies in \autoref{tab:app1}.
	\begin{table}[htb!]
		\centering
            \renewcommand{\arraystretch}{1.2}{
		\begin{tabular}{|c|c|c|}
			\hline
			$\sqrt{s}$ (GeV) & $\sigma_{\chi \overline{\chi} \gamma}$ (fb) & $\sigma_{\nu \overline{\nu} \gamma}$ (fb) \\ \hline
			250 & 0.0022 & 2743 \\
			500 & 0.0358 & 2120 \\
			1000 & 0.5889 & 2495 \\ \hline
		\end{tabular}}
		\caption{Production cross sections for signal ($\chi \overline{\chi} \gamma$, {\color{black}for the benchmark: $m_{\chi}$=33 MeV and $\Lambda$=1.14 TeV}) and dominant background ($\nu \overline{\nu} \gamma$) at different CM energies of the ILC.}
		\label{tab:app1}
	\end{table}
	The simulated events are showered in \texttt{Pythia8} to replicate ISR and FSR effects. The showered events are fed to \texttt{Delphes3} for detector simulation. The details of particle identification and isolation criteria used in \texttt{Delphes3} are listed below:
	\begin{itemize}
		\item \textbf{Photon}: Identification requires $p^{\gamma}_{T} > 10$ GeV and $|\eta_{\gamma}| < 2.5$. The identification efficiency for $|\eta_{\gamma}| \le 1.5$ and $1.5 < |\eta_{\gamma}| \le 2.5$ regions are 0.95 and 0.85 respectively. The photon isolation cone is taken to be $\Delta R = 0.5$ and the photon is isolated if the isolation $p_{T}$ ratio is less than 0.12\footnote{The $p_{T}$ isolation ratio is the sum of the $p_{T}$ of all other species in the $\Delta R$ cone to the $p_{T}$ of the photon.}.
		
		\item \textbf{Lepton}: Identification requires $p^{\ell}_{T} > 10 \,(10)$ GeV and $|\eta_{\ell}| < 2.5 \,(2.4)$ for electron (muon). The identification efficiency for $|\eta_{\gamma}| \le 1.5$ and $1.5 < |\eta_{\gamma}| \le 2.5\, (2.4)$ regions for electrons (muons) are 0.95 (0.95) and 0.85 (0.95) respectively. The electron (muon) isolation cone is taken to be $\Delta R = 0.5 \,(0.5)$ and the electron (muon) is isolated if the isolation $p_{T}$ ratio is less than 0.12 (0.25).
		
		\item \textbf{Jet}: The jet clustering is done using the anti-$k_{T}$ algorithm with jet radius, $R = 0.5$ and $p^{j}_{T} > 20$ GeV.
	\end{itemize}
	The tracking efficiencies and momentum resolution functions are same as the default \texttt{Delphes3} card. These detector efficiencies combined consitute the `Basic Cuts' in Tab. I of the letter. As discussed in the letter, we additionally implement three more sequential cuts: $\slashed{E}_{T} > 200$ GeV, $\slashed{E} \in [525, 750]$ GeV, and $|\eta_{\gamma}| > 1.0$. The detailed cutflow of the signal and the dominant background for different polarization combination are detailed in \autoref{tab:app2} with associated Poissonian uncertainties. The benchmark is same as \autoref{tab:sbc}.
    \begin{table*}[htb!]
	\centering
        \renewcommand{\arraystretch}{1.2}{
	\begin{tabular}{|cc|c|c|c|c|c|}
		\hline
		\multicolumn{2}{|c|}{} & \multicolumn{5}{c|}{Polarization, $\{P_{e^{+}}, P_{e^{-}}\}$ } \\ \cline{3-7}
			 \multicolumn{2}{|c|}{Cutflow} & $\{0 \%, 0 \%\}$ & $\{+20 \%, +80 \%\}$ & $\{-20 \%, +80 \%\}$ & $\{+20 \%, -80 \%\}$ & $\{-20 \%, -80 \%\}$ \\ \hline
			 \multirow{2}{*}{Basic Cuts} & (S) & 4422 $\pm$ 66 & 3710 $\pm$ 61 & 5124 $\pm$ 72 & 5124 $\pm$ 72 & 3701 $\pm$ 61 \\ \cline{3-7}
			 & (B) & 18101325 $\pm$ 4254  & 4544778 $\pm$ 2132 & 3354590 $\pm$ 1832 & 37339851 $\pm$ 6111 & 24934392 $\pm$ 4993 \\ \hline
			 \multirow{2}{*}{$\slashed{E}_{T} > 200$ GeV} & (S) & 2957 $\pm$ 54 & 2484 $\pm$ 50 & 3419 $\pm$ 58 & 3434 $\pm$ 59 & 2479 $\pm$ 50 \\ \cline{3-7}
			 & (B) & 725945 $\pm$ 852  & 344134 $\pm$ 587 & 413257 $\pm$ 643 & 1193090 $\pm$ 1092 & 831588 $\pm$ 912 \\ \hline
			 \multirow{2}{*}{$\slashed{E} \in [525, 750]$ GeV} & (S) & 2694 $\pm$ 51 & 2264 $\pm$ 48 & 3115 $\pm$ 56 & 3129 $\pm$ 56 & 2256 $\pm$ 47 \\ \cline{3-7}
			& (B) & 289420 $\pm$ 538  & 73299 $\pm$ 271 & 54545 $\pm$ 234 & 588283 $\pm$ 767 & 395508 $\pm$ 629 \\ \hline
			 \multirow{2}{*}{$|\eta_{\gamma}| > 1.0$} & (S) & 2395 $\pm$ 49 & 2010 $\pm$ 45 & 2778 $\pm$ 53 & 2774 $\pm$ 53 & 2003 $\pm$ 45 \\ \cline{3-7}
			& (B) & 219161 $\pm$ 468  & 55052 $\pm$ 235 & 40711 $\pm$ 202 & 453639 $\pm$ 674 & 303079 $\pm$ 550 \\ \hline
	\end{tabular}}
	\caption{Signal ($S$), $\chi \overline{\chi} \gamma$ and dominant background ($B$), $\nu \overline{\nu} \gamma$ event counts with uncertainties in each step of the subsequent cuts for mono-$\gamma$ final state signal {\color{black}(for the benchmark: $m_{\chi}$=33 MeV and $\Lambda$=1.14 TeV)} for different polarization combinations at the 1 TeV ILC with $\mathfrak{L}_{\rm int}=8$ ab$^{-1}$ and different possible beam polarizations.}
	\label{tab:app2}
    \end{table*}
	
\paragraph{\textbf{Other Background Processes}} Other possible backgrounds arise from processes like $e^{+} e^{-} \rightarrow W^{+} W^{-} \gamma$, $e^{+} e^{-} \rightarrow ZZ\gamma$, $e^{+} e^{-} \rightarrow \ell^{+} \ell^{-} \gamma$, and $e^{+} e^{-} \rightarrow \gamma \gamma \gamma$, where the outgoing particles or decay products are undetected, resulting in missing energy. In \autoref{tab:app3}, we provide an estimate on the backgrounds $W^{+}W^{-}\gamma$ and $ZZ\gamma$ for the cutflow in \autoref{tab:app2}, and gauge their effect on the signal significance. We only show the numbers of the unpolarized case and the optimal polarization choice of $\{P_{e^{+}}, P_{e^{-}}\} = \{-20 \%, +80 \%\}$. The backgrounds $\ell^{+}\ell^{-}\gamma$ and $\gamma \gamma \gamma$ do not survive till the final cut and hence not listed. From the last two rows of \autoref{tab:app3}, it is clear that the signal significance do not alter significantly and hence considering $\nu \overline{\nu} \gamma$ is good enough to obtain an estimate on the discovery limit.
\begin{table}[htb!]
    \centering
    \renewcommand{\arraystretch}{1.2}{
    {\small
    \begin{tabular}{|c|c|c|c|}
    \hline
    \multicolumn{2}{|c|}{} & \multicolumn{2}{c|}{Polarization, $\{P_{e^{+}}, P_{e^{-}}\}$ } \\ \cline{3-4}
	\multicolumn{2}{|c|}{Cutflow} & $\{0 \%, 0 \%\}$ & $\{-20 \%, +80 \%\}$  \\ \hline
	\multirow{2}{*}{Basic Cuts} & $W^{+}W^{-}\gamma$ & 2136 $\pm$ 46 & 5814 $\pm$ 76 \\ \cline{3-4}
	& $ZZ\gamma$ &  3805 $\pm$ 62  & 5429 $\pm$ 74 \\ \hline
	\multirow{2}{*}{$\slashed{E}_{T} > 200.0$ GeV} & $W^{+}W^{-}\gamma$ & 334 $\pm$ 18 & 1024 $\pm$ 32 \\ \cline{3-4}
	& $ZZ\gamma$ & 522 $\pm$ 23  & 762 $\pm$ 28 \\ \hline
	\multirow{2}{*}{$\slashed{E} \in [525.0, 750.0]$ GeV} & $W^{+}W^{-}\gamma$ & 242 $\pm$ 16 & 738 $\pm$ 27 \\ \cline{3-4}
	& $ZZ\gamma$ & 468 $\pm$ 22  & 676 $\pm$ 26  \\ \hline
	\multirow{2}{*}{$|\eta_{\gamma}| > 1.0$} & $W^{+}W^{-}\gamma$ & 175 $\pm$ 13 & 504 $\pm$ 22  \\ \cline{3-4}
	& $ZZ\gamma$ & 348 $\pm$ 19  & 530 $\pm$ 23  \\ \hline
	\multicolumn{2}{|c|}{Significance (only $\nu \nu \gamma$)} & 5.11 $\pm$ 0.10 & 13.62 $\pm$ 0.26 \\ \hline
	\multicolumn{2}{|c|}{Significance (all backgrounds)} & 5.10 $\pm$ 0.10 & 13.45 $\pm$ 0.26 \\ \hline
    \end{tabular}}}
    \caption{Event counts of backgrounds $W^{+}W^{-}\gamma$ and $ZZ\gamma$ with uncertainties in each step of the subsequent cuts for mono-$\gamma$ final state signal {\color{black}(for the benchmark: $m_{\chi}$=33 MeV and $\Lambda$=1.14 TeV)} and the final signal significance in presence and absence of these backgrounds at the 1 TeV ILC with $\mathfrak{L}_{\rm int}=8$ ab$^{-1}$ and two different beam polarizations.}
\label{tab:app3}
\end{table}

\section{Recasting Limits from the LEP}
\label{app:LEP}
\begin{figure}[htb!]
\centering    
\includegraphics[width = 0.95\linewidth]{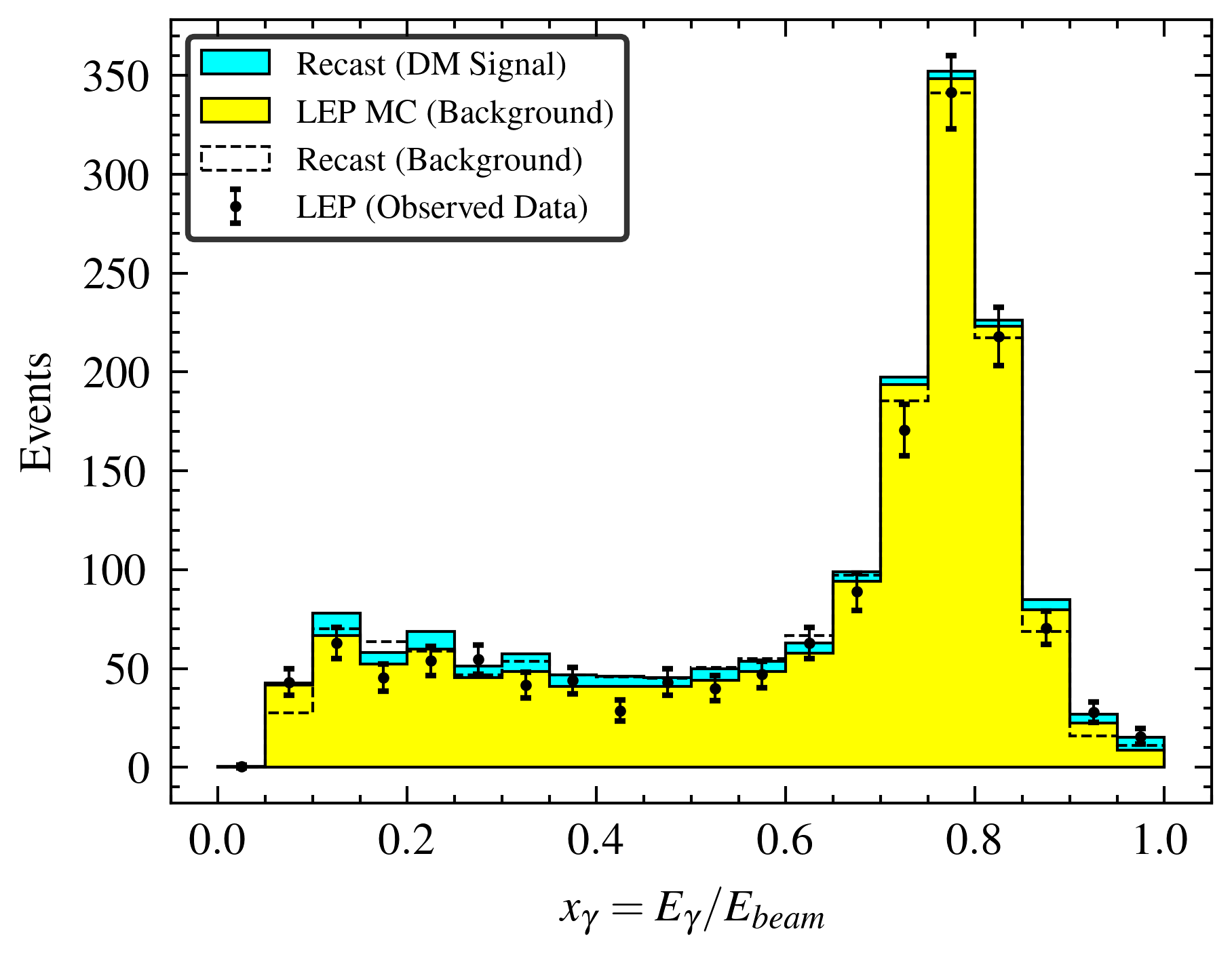}
\includegraphics[width = 0.95\linewidth]{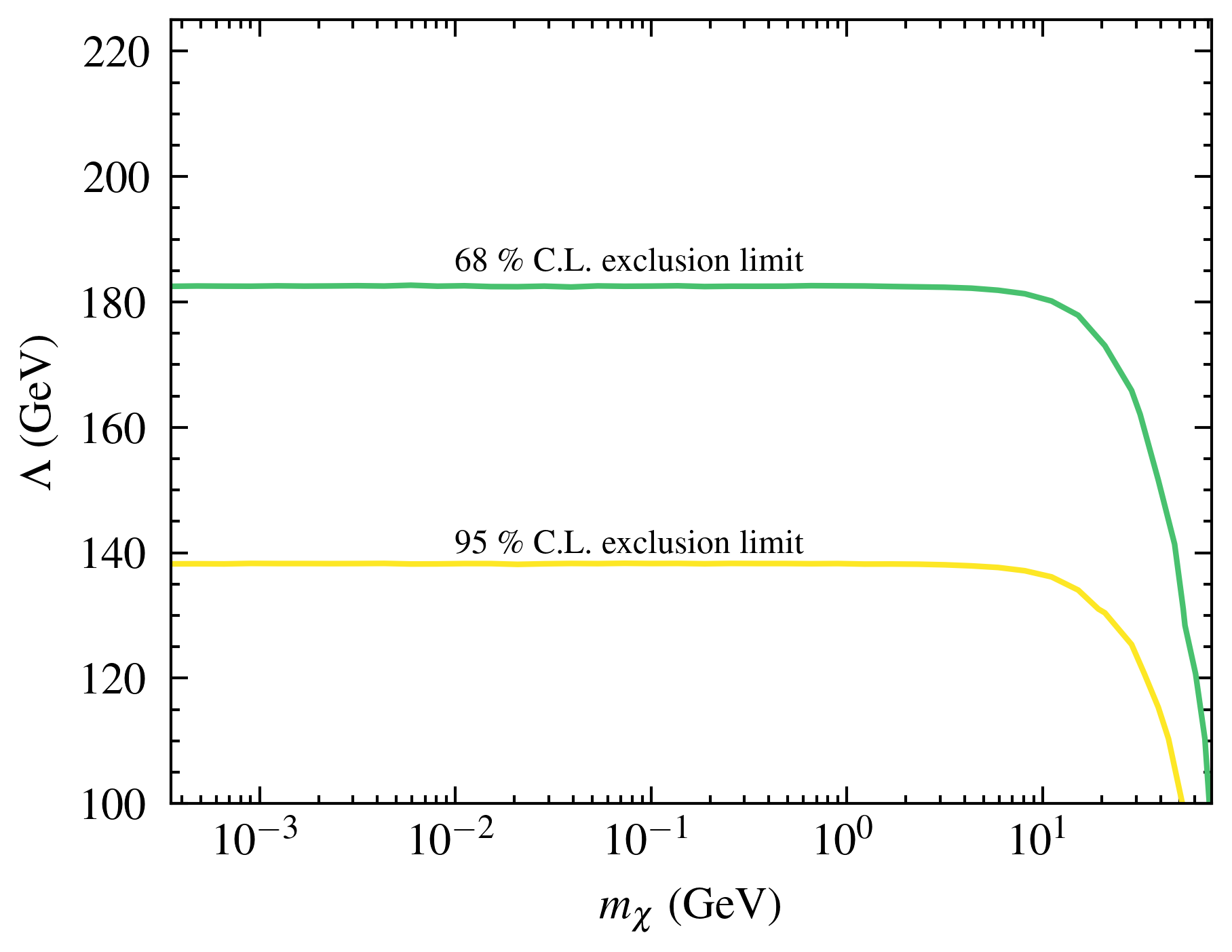}
\caption{{\it Left:} Binned $x_{\gamma}$ distribution for LEP studies corresponding to $\mathfrak{L}_{int}$ = 650 pb$^{-1}$. The DM signal corresponds to $\chi \overline{\chi} \gamma$ production for the benchmark $m_{\chi} = 1.37$ MeV, $\Lambda = 0.23$ TeV. {\it Right:} 68\% C.L. and 95\% C.L exclusion limits from LEP recast study on the $m_{\chi}-\Lambda$ plane.}
\label{fig:lep}
\end{figure}

The Large Electron–Positron Collider (LEP) has performed a number of studies concerning single photon events with missing energy~\cite{DELPHI:2003dlq, DELPHI:2008uka}. However, most of them were aimed towards precision measurements within the SM or other exotic searches. One way to extract information from these existing studies is to recast them using identical set of resolutions and efficiencies as used in the original studies and feed Monte Carlo (MC) generated events through the recasting framework. Such study has been done concerning the leptophilic operators in, for example, Ref.~\cite{Fox:2011fx}. Here we provide an exclusion bound on the $m_{\chi}-\Lambda$ plane based on existing LEP data from mono-$\gamma$ studies. LEP studies were done over a range of CM energies $\sqrt{s}=[180-209]$ GeV. The observed events, however, were reconstructed and presented in bins of $x_{\gamma}$, defined as $E_{\gamma} / E_{\rm beam}$, thus independent of CM energy of the collisions. For our analysis, we generate events at a reference CM mass energy of $\sqrt{s} = 200$ GeV, which, following the conclusion drawn in~\cite{Fox:2011fx}, does not invalidate the LEP data. The DELPHI detector at LEP had three angular regions (HPC, FEMC and STIC), and each region had a different set of trigger and reconstruction/identification efficiencies. The recasting of each of these regions, based on~\cite{DELPHI:2003dlq, DELPHI:2008uka}, are detailed in \autoref{tab:recast}. There is an additional photon identification efficiency of 90\% valid for all regions. For STIC, since, the information provided by~\cite{DELPHI:2003dlq} is incomplete, we assume the overall efficiency to be 48\% as done by \cite{Fox:2011fx}. In order to validate our framework, we plot the $x_{\gamma}$ distribution of our generated MC events for the SM backgrounds on top of the same from DELPHI MC studies. They are found to be in perfect agreement as shown in \autoref{fig:lep}. The observed data as well as the DM signal corresponding to the benchmark $m_{\chi} = 1.37$ MeV, $\Lambda = 0.23$ TeV.
\begin{table*}[htb!]
\centering
\renewcommand{\arraystretch}{1.2}{
\begin{tabular}{|c|c|c|c|}
\hline
Detector  & Trigger & Reconstruction & Energy \\
Regions & Efficiency & Efficiency & Smearing \\ \hline
& $x_{\gamma} \in [0.06, 0.30]$ &  $x_{\gamma} \in [0.06, 0.80]$ & Gaussian ($\sigma_{E}/E_{\gamma}$): \\
HPC & (45.75 + 1.042 $E_{\gamma}$)\% &  (38 + 0.5 $E_{\gamma}$)\% & $ 0.043 \oplus 0.32/ \sqrt{E_{\gamma}} $ \\
$\theta_{\gamma} \in [90^{\circ}, 45^{\circ}]$  & $\cup$ & $\cup$ & + \\
$x_{\gamma} \in [0.06, 1.10]$  & $x_{\gamma} \in [0.30, 1.10]$ &$x_{\gamma} \in [0.80, 1.10]$  & Lorentzian ($\Gamma$): \\
& (74 + 0.1 $E_{\gamma}$)\% & 78\% & 0.04 $E_{\gamma}$ \\
\hline
& $x_{\gamma} \in [0.10, 0.15]$ &  & Gaussian ($\sigma_{E}/E_{\gamma}$): \\
FEMC & (79 + 1.4 $E_{\gamma}$)\% &  & $ 0.03 \oplus 0.12/ \sqrt{E_{\gamma}} \oplus 0.11/ E_{\gamma} $  \\
$\theta_{\gamma} \in [32^{\circ}, 12^{\circ}]$  & $\cup$ & $0.89(55 + 0.2 E_{\gamma})\%$ & + \\
$x_{\gamma} \in [0.10, 0.90]$  & $x_{\gamma} \in [0.15, 0.90]$ &  & Lorentzian ($\Gamma$): \\
& 100\% &  & 0.02 $E_{\gamma}$ \\
\hline
& \multicolumn{2}{c|}{} & Gaussian ($\sigma_{E}/E_{\gamma}$): \\
STIC & \multicolumn{2}{c|}{} & $0.0152 \oplus 0.135/ \sqrt{E_{\gamma}}$ \\
$\theta_{\gamma} \in [8^{\circ}, 3.8^{\circ}]$  & \multicolumn{2}{c|}{48\%} & + \\
$x_{\gamma} \in [0.30, 0.90]$  & \multicolumn{2}{c|}{} & Lorentzian ($\Gamma$):  \\
& \multicolumn{2}{c|}{} & 0.02 $E_{\gamma}$ \\
\hline
\end{tabular}}
\caption{Details of resolution and efficiencies for the LEP recast study. The $\theta_{\gamma}$ ranges are shown for one half of the detector only, but the same efficiencies apply for the other half as well. Additionally we implement angular cuts: $\theta_{\gamma} > (28 - 80 x_{\gamma})^{\circ}$ and $\theta_{\gamma} > (9.2 - 9 x_{\gamma})^{\circ}$ for FEMC and STIC regions respectively. For further details, see Ref.~\cite{Fox:2011fx}.}
\label{tab:recast}
\end{table*}
To obtain an exclusion bound on the $m_{\chi}-\Lambda$ plane, we perform a $\Delta \chi^{2}$ test for the binned $x_{\gamma}$ distribution. We exclude the first bin and consider other 19 bins for the analysis. The $\Delta \chi^{2}$ for this case is defined as:
\begin{equation}
\Delta \chi^{2} = \sum^{19}_{i = 1} \left(\frac{N_{\rm obs} - (S (m_{\chi}, \Lambda) + B)}{\sqrt{S(m_{\chi}, \Lambda) + B}}\right)^{2},
\end{equation}
where, $N_{\rm obs}$ is the number of events observed in each bin, $S$ and $B$ are number of signal and background events post detector efficiencies. The degrees of freedom (dof) for the binned $\Delta \chi^{2}$ analysis is $N - M = 19 - 2 = 17$, where, $N$ is the number of bins and $M$ is the number of model parameters. For 17 dofs, the $\Delta \chi^{2}$ value corresponding to 68\% C.L. and 95\% C.L. are 19.514 and 27.587 respectively. \autoref{fig:lep} shows, for low mass regime, the $\Lambda$ cutoff is around 180 GeV and 140 GeV at 68\% C.L. and 95\% C.L. respectively.

\section{Recasting Limits from the LHC}
\label{app:LHC}
In the previous segment, we successfully recast the LEP analysis to explore its implications for our model. Building on that foundation, our focus now shifts to recasting specific LHC studies that investigate particle production processes associated with missing transverse energy (MET). By leveraging existing analyses, we aim to establish robust constraints on our model, further refining its compatibility with current experimental data. Within this recast framework, we further provide projected $95\%$ C.L. exclusion limits for the HL-LHC.
\begin{figure}[htb!]
\centering
\begin{tikzpicture}
\begin{feynman}
\vertex(a);
\vertex[above left =1cm and 0.5cm of a] (a1){$q$};
\vertex[below left =1cm and 0.5cm of a] (a2){$\overline{q}$};
\vertex[right = 1cm of a] (c);
\vertex[above right =1cm and 0.5cm of c] (b1){\({\color{black}\rm \chi} \)};
\vertex[below right =1cm and 0.5cm of c] (b2){\({\color{black}\rm \overline{\chi}}\)};
\vertex[right = 0.75cm of c] (b3){\({\color{black}\rm\gamma}\)};
\diagram*{
(a1) -- [ fermion, arrow size=0.7pt,edge label={\(\rm \)},style=black] (a),
(a) -- [ fermion, arrow size=0.7pt,edge label={\(\rm \)},style=black] (a2),
(a) -- [ boson, edge label={\(\rm{\color{black} \gamma} \)},style=gray!75] (c),
(c) -- [fermion, arrow size=0.7pt,edge label={\(\rm \)},style=red,edge label={\({\color{black}} \)}] (b1),
(b2) -- [fermion, arrow size=0.7pt,style=red,edge label={\({\color{black}\rm} \)}] (c),
(b3) -- [boson ,style=red,edge label={\({\color{black}\rm} \)}] (c)};
\end{feynman}
\node at (c)[black,fill,style=black,inner sep=2pt]{};
\end{tikzpicture}
\hspace{1cm}
\begin{tikzpicture}
\begin{feynman}
\vertex(a);
\vertex[above left =1cm and 0.5cm of a] (a1){$q$};
\vertex[below left =1cm and 0.5cm of a] (a2){$\overline{q}$};
\vertex[right = 1cm of a] (c);
\vertex[above right =1cm and 0.5cm of c] (b1){\({\color{black}\rm \chi} \)};
\vertex[below right =1cm and 0.5cm of c] (b2){\({\color{black}\rm \overline{\chi}}\)};
\vertex[right = 0.75cm of c] (b3){\({\color{black} Z}\)};
\diagram*{
(a1) -- [ fermion, arrow size=0.7pt,edge label={\(\rm \)},style=black] (a),
(a) -- [ fermion, arrow size=0.7pt,edge label={\(\rm \)},style=black] (a2),
(a) -- [ boson, edge label={\(\color{black} Z \)},style=gray!75] (c),
(c) -- [fermion, arrow size=0.7pt,edge label={\(\rm \)},style=red,edge label={\({\color{black}} \)}] (b1),
(b2) -- [fermion, arrow size=0.7pt,style=red,edge label={\({\color{black}\rm} \)}] (c),
(b3) -- [boson ,style=red,edge label={\({\color{black}\rm} \)}] (c)};
\end{feynman}
\node at (c)[black,fill,style=black,inner sep=2pt]{};
\end{tikzpicture}
\caption{Feynman diagrams of mono-$\gamma$ (\textit{left}) and dilepton + MET (via $Z$ decay to leptons, \textit{right}) signals at the LHC.}
\label{fig:ldm-lhc}
\end{figure}

\paragraph{\textbf{Recasting LHC 13 TeV analyses}}We consider two possible DM signal processes in the context of the LHC viz.  mono-$\gamma$ and dilepton + MET (mono-$Z$ production followed by decay to leptons) shown in \autoref{fig:ldm-lhc}. There have been searches at the LHC regarding these signal channels in the context of different models. In order to obtain a limit on our parameter space from the experimental results, we recast the experimental analysis in the context of our model. 
\begin{figure}[htb!]
\centering    
\includegraphics[width=0.95\linewidth]{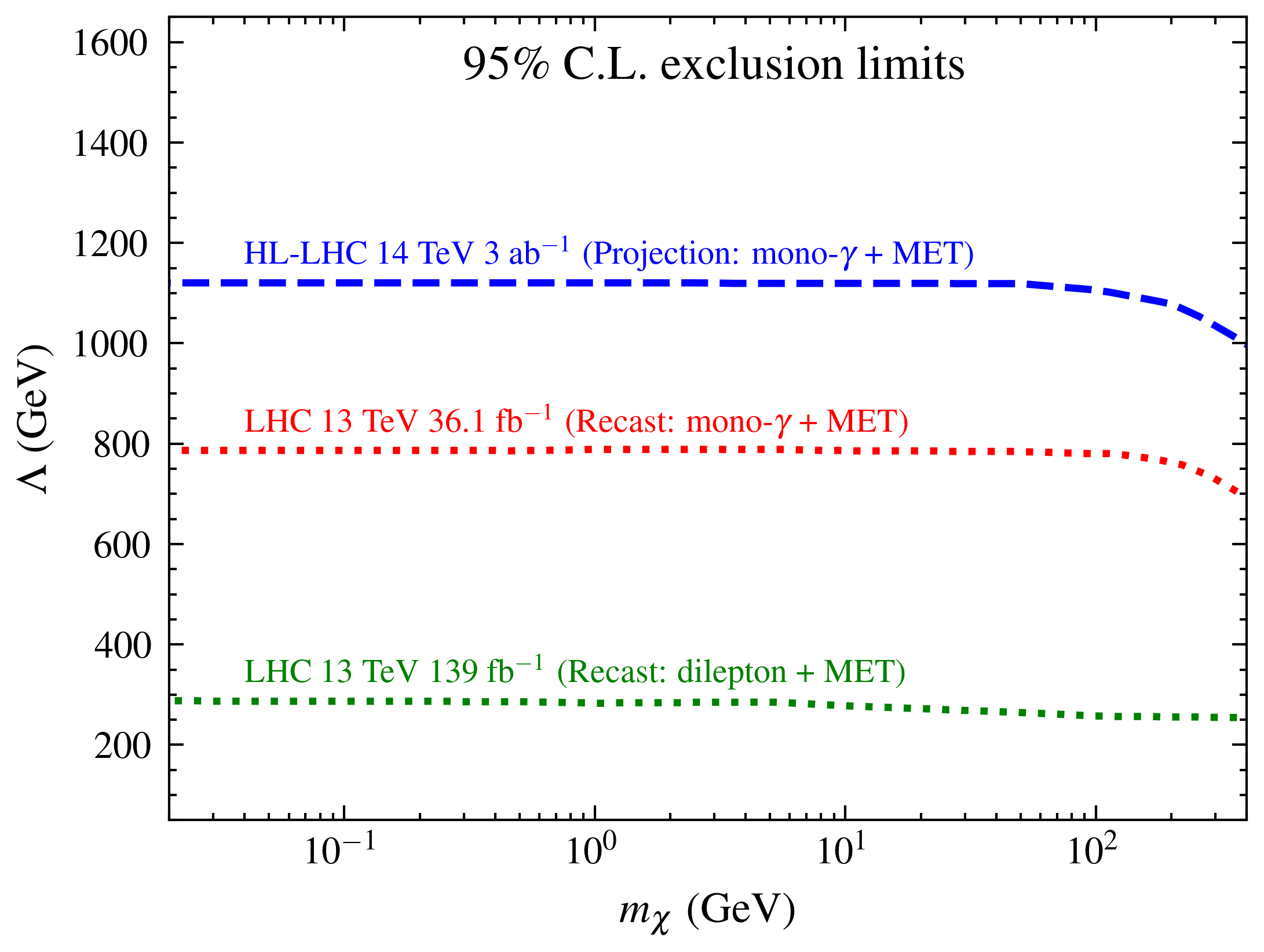}
\caption{95\% C.L. exclusion limits from the LHC recast and HL-LHC projection study on the $m_{\chi}-\Lambda$ plane.}
\label{fig:lhc}
\end{figure}
This is done using \texttt{CheckMATE2} \cite{Dercks:2016npn}, which is based on \texttt{Delphes3} and uses the CL$_{S}$ method \cite{Read:2002hq} to provide 95\% C.L. exclusion limits on the parameter spaces of the models. For the mono-$\gamma$ process i.e. $pp \rightarrow \chi \overline{\chi} \gamma$, we recast the ATLAS analysis \cite{ATLAS:2017nga}, which search for DM in final states containing an energetic photon and large missing transverse momentum at $\sqrt{s} = 13$ TeV (at an integrated luminosity of 36.1 fb$^{-1}$) using \texttt{atlas\_1704\_03848} analysis of \texttt{CheckMATE2}. The signal regions are same as \cite{ATLAS:2017nga}. The event processes is generated at different benchmarks throughout the $m_{\chi}-\Lambda$ parameter space in \texttt{MG5\_aMC} and is showered in \texttt{Pythia8}. The showered events are then fed to \texttt{CheckMATE2} which recasts the experimental analysis results using custom \texttt{Delphes3} cards for different signal regions to provide the exclusion limits. For the mono-$\gamma$ signal, $\Lambda < 800$ GeV is excluded by the LHC recast for the low mass regime. A similar analysis is done for dilepton + MET signal arising from $pp \rightarrow \chi \bar{\chi} Z(\ell^+ \ell^-) $. We consider two possible DM signal processes in the context of the LHC viz.  mono-$\gamma$ and dilepton + MET (mono-$Z$ production followed by decay to leptons) shown in \autoref{fig:ldm-lhc}. There have been searches at the LHC regarding these signal channels in the context of different models. In order to obtain a limit on our parameter space from the experimental results, we recast the experimental analysis in the context of our model. This is done using \texttt{CheckMATE2} \cite{Dercks:2016npn}, which is based on \texttt{Delphes3} and uses the CL$_{S}$ method \cite{Read:2002hq} to provide 95\% C.L. exclusion limits on the parameter spaces of the models. For the mono-$\gamma$ process i.e. $pp \rightarrow \chi \overline{\chi} \gamma$, we recast the ATLAS analysis \cite{ATLAS:2017nga}, which search for DM in final states containing an energetic photon and large missing transverse momentum at $\sqrt{s} = 13$ TeV (at an integrated luminosity of 36.1 fb$^{-1}$) using the \texttt{atlas\_1704\_03848} analysis of \texttt{CheckMATE2}. The signal regions are same as \cite{ATLAS:2017nga}. The event processes are generated at different benchmarks throughout the $m_{\chi}-\Lambda$ parameter space in \texttt{MG5\_aMC} and are showered in \texttt{Pythia8}. The showered events are then fed to \texttt{CheckMATE2} which recasts the experimental analysis results using custom \texttt{Delphes3} cards for different signal regions to provide the exclusion limits, shown in \autoref{fig:ldm-lhc}. For the mono-$\gamma$ signal, $\Lambda < 800$ GeV is excluded by the LHC recast for the low mass regime. A similar analysis is done for dilepton + MET signal arising from the process $pp \rightarrow \chi \overline{\chi} Z (\ell^{+} \ell^{-})$. We recast the ATLAS analysis \cite{ATLAS:2019lff} at $\sqrt{s} = 13$ TeV (at an integrated luminosity of 139 fb$^{-1}$) using the \texttt{atlas\_1908\_08215} analysis of \texttt{CheckMATE2}. The exclusion limit is less stringent compared to mono-$\gamma$ case and is also shown in \autoref{fig:ldm-lhc}. We conclude that our analysis regime $i.e.$ $\Lambda > 1$ TeV is safe from the bounds from the 
recast of experimental results of the LHC experiments.

\paragraph{\textbf{Projecting HL-LHC limits}} To project exclusion limits for the future HL-LHC run at a CM energy of 14 TeV with an integrated luminosity of 3 ab$^{-1}$, we adopt the same signal regions (SRs) as those defined in Ref.~\cite{ATLAS:2017nga}. Background event yields in each SR are estimated for 14 TeV and 3 ab$^{-1}$ using \texttt{MG5\_aMC}, in conjunction with the \texttt{atlas\_1704\_03848} analysis card implemented in \texttt{CheckMATE2}. The expected 95\% C.L. signal upper limit, $S_{95}$, in each SR is scaled by a factor of $\sqrt{R_B}$, where $R_B$ is the ratio of background event yields at 14 TeV with 3 ab$^{-1}$ to those at 13 TeV with 36.1 fb$^{-1}$. The exclusion is quantified using the parameter $r$, defined as: $r=(S - 1.64 \cdot \Delta S)/S_{95}$, where $S$ and $\Delta S$ denote the predicted signal yield and its associated uncertainty for a given parameter point in a specific SR. A value of $r > 1$ indicates exclusion at 95\% C.L. The projected HL-LHC exclusion reach is presented in \autoref{fig:lhc}. In the low-mass regime, the 95\% C.L. exclusion limit extends up to $\Lambda \sim 1.1$ TeV.
{\color{black}
\section{Validity of the EFT approach}
\label{app:EFT}
The effective theory realization of a New Physics (NP) scenario relies on a perturbative expansion of the production cross section in inverse powers of the new physics scale, $\Lambda$. At the amplitude level, the EFT contribution can be expressed as
\begin{equation}
\mathcal{A} \;=\; \sum_{d} \, \widetilde{A}_{d} 
\left(\dfrac{p}{\Lambda}\right)^{d-4},
\end{equation}
where $d$ denotes the canonical dimension of the effective operator and $p$ represents the typical momentum scale of the process. The perturbative expansion is well-defined only if the expansion parameter is sufficiently small, i.e.
\begin{equation}
p < \Lambda,
\end{equation}
ensuring the convergence of the series and thereby the validity of the EFT description~\cite{Manohar:2018aog}.

A general requirement for EFT validity is that the cutoff scale should lie above the largest energy scale probed in the process under consideration. In the context of the early Universe, where DM production is governed by the thermal bath after reheating, the highest relevant energy scale is the reheating temperature $T_{\rm RH}$. The EFT description remains valid only if
\begin{equation}
\Lambda > T_{\rm RH} \gtrsim m_{\rm DM},
\end{equation}
ensuring that the mediator responsible for DM interactions is not kinematically accessible and that the EFT consistently captures the dynamics of DM production without reference to the details of the UV completion. 

At collider experiments, the same principle applies: the EFT expansion is justified only if the accessible momenta remain below the cutoff scale. For lepton colliders, the maximum momentum transfer is set by the CM energy, $p \sim \sqrt{s}$. Thus, the validity condition reduces to
\begin{equation}
\sqrt{s} < \Lambda.
\end{equation}
In the case of the ILC, where we consider $\sqrt{s} = 1$ TeV, the EFT analysis is self-consistent only if the new physics scale satisfies $\Lambda > 1$ TeV. This guarantees that higher-dimensional operator contributions remain perturbative, while the underlying UV completion remains safely decoupled at the energies probed by the collider. 
}
\bibliographystyle{elsarticle-num} 
\bibliography{ldm}

\end{document}